# Elongated element as a model of point contact in ion-conducting medium

A.P. Pospelov[1], G.V. Kamarchuk[2]

[1]National Technical University "Kharkov Polytechnical Institute", 61002 Kharkov, Ukraine
[2]B.Verkin Institute for Low Temperature Physics & Engineering of NAS of Ukraine, 61103 Kharkov, Ukraine

## Abstract

In present work it is suggested the concept of continual multi-electrode system being realized on the elongated electrically conducting element immersed into electrolyte. Conjugation of such kind of system with the point-contact structure creates the unique electrode architecture and permits realizing of the auto-oscillation process of electrochemical commutation at atomic scale. During this process the nanostructure is being realized in a form of point contact at the place of the mutual touching of the dendrite top and counter-electrode. The ability of point contacts to concentrate the electric field in assemblage with the peculiar properties of elongated element causes unambiguously the localization of all the electrochemical transformations at the dendrite top, where contact is forming.

The point contacts are taking special place among other known nanostructures. They have gained the great popularity well before the appearance of the modern widely spread investigations of nanosized objects due to the usage of point contacts as a powerful tool for studying of fundamental physical properties of materials, for example, of the electron-phonon interaction in metals [1]. Traditionally these investigations were carried out at the liquid helium temperature, permitting minimization of the influence of thermal vibrations of the crystal lattice on the character of spectra of the electron-phonon interaction. Those experiments were based on precise study of the current-voltage characteristic of the point contact, caused by physical properties of material, which has formed the bulk part of the conductivity channel. Novel investigations, which were carried out with accounting of the surface phenomena at room temperature, have allowed discovering the effect of the enhanced sensory ability of point contacts [2]. This circumstance has become the start moment toward real practical applications of point contacts and, at the same time, the beginning of research toward improvement and optimization of methods of their obtaining. One of the prospective directions of solving these problems is electrochemical growth of metallic dendrites and forming of the point contacts on dendrite basis [3, 4].

In fundamental low-temperature investigations the point contacts are being created and used in non-conducting medium. On the other hand electrochemical synthesis of point contacts for sensory applications occur in a medium of ionic conductor, requiring development of the new models, which should ascribe the properties of such kind of objects. Let us consider in this article the concept of the elongated element [5, 6] in order to show the expediency of implementation of this concept for characterization and explanation of nanostructural effects in the system containing the point contact in the ion-conducting surrounding.

In majority of cases electrochemical systems consist of electrodes segregated by electrolyte. Classical electrochemical circuit with external source has series placing of ionic and electronic conductors [7, 8]. If some reasons cause existence of electronic conductor parallel to ionic part of the circuit, represented by electrolyte, than practically all current flows through the former because specific resistance of electronic conductor is by several orders of magnitude lower than that of electrolyte, as well as of the polarization resistance of electrochemical reactions. Electrochemical circuit reveals itself shunted one, and realization of desired electrochemical reactions becomes inexpedient.

All the above statements are valid for macroscopic objects of consideration. However if the system under consideration contains certain nanostructures, the local frequencies of the electrode processes may reveal themselves enough for causing essential effects even in presence of electronic conductor included in parallel in the schematic. Behavior of such kind of the electrode system in electric field is described with the help of equivalent schematic, which represents itself as electric circuit consisting of included in a certain manner various electric engineering elements. For synthesis of such kind of schematic one should fulfill measurements of the Ohm (just resistive) and capacitive components of the input impedance of the electrode systems [9] with applied sinusoidal voltage of minor (up to 5 mV) amplitude.

In general case, electric model of the system, consisting of two electrodes segregated by the electrolyte layer, may be represented by the equivalent schematic shown in Fig. 1. Here the active resistance $R_e$ is determined as a value inversed to the conductivity of segregating electrolyte layer; C is capacitance of the double electric layer, which is being formed at the «electrode – ion-conducting medium» interface; $Z_R$ is an impedance characterizing the degree of deceleration of electronic transition at the interface boundary, as well as of the processes of adsorption of the electrolyte components and ionic transport, and also the mutual interaction of these processes.

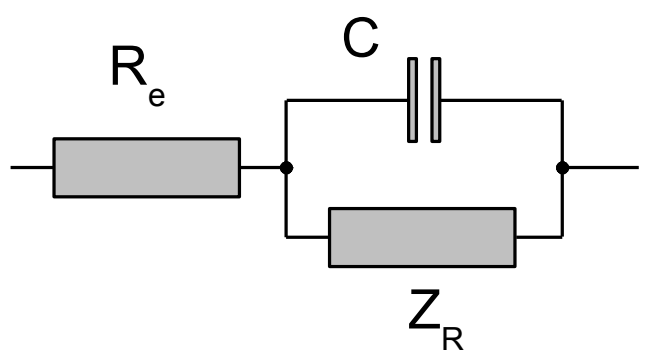

**Figure 1.** Generalized equivalent schematic of the double-electrode system with the segregating electrolyte layer.

If the present system electrodes are connected with a help of metallic conductor (for instance, by the dendrite, which has grown in a process of electrochemical precipitation of metal ions), than the equivalent schematic is being complicated a bit, namely, the scheme of Fig. 1 should be completed by two additional parallel circuits according to two new trajectories of the charge carriers. In the latter case an option of the charge transfer through metallic conductor with an active resistance $R_m$ appears together with initial scenario, according to which electric current flows via the surfaces of electrodes and electrolyte. As the latter conductor is placed in electric field and contacts with electrolyte, one may consider it in a role of so-called elongated element [5, 6], at which surface electrochemical reactions occur under certain conditions. General view of equivalent schematic is presented in Fig. 2. Here $R_e$, $C$ and $Z_R$ are parameters of the output electrode system, which were described earlier (assuming negligibly small influence of the screening effect of metallic conductor, which connects electrodes); $R_m$ is an active resistance of metallic conductor, shunting the electrode system; $R^*_e$, $C^*$ and $Z^*_R$ and integral values of parameters of the electrode system, being realized on a basis of the elongated element, namely: the electrolyte resistance, capacity of the double electric layer and impedance characterizing the final frequency of the electrode processes, respectively.

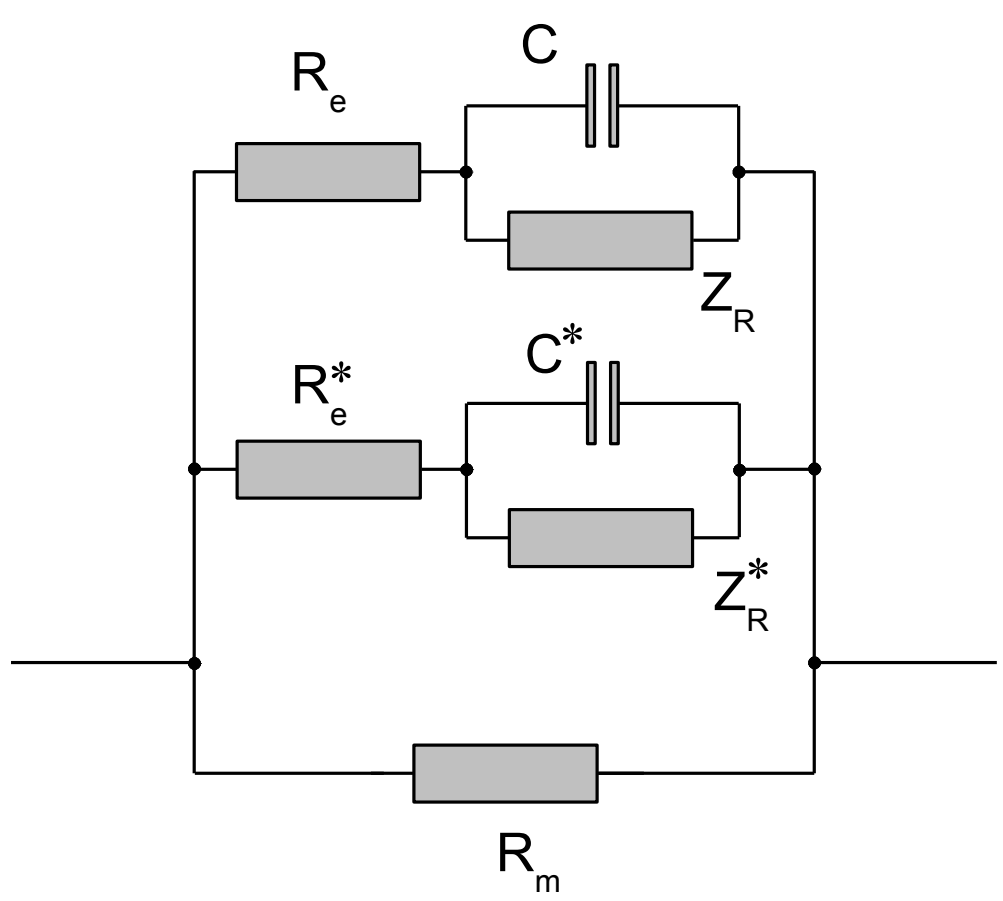

**Figure 2.** Generalized equivalent schematic of the system of two electrodes connected by electronic conductor.

The voltage applied to the ends of elongated element is distributed along its longitudinal axis in a certain fashion. The character of this distribution is determined, first of all, by the conductor form and its nature. Taking this circumstance into account, the values of $R^*_e$, $C^*$ and $Z^*_R$ can be obtained by means of integrating of respective parameters along the conductor with accounting of the potential distribution. The value and sign of potential in every point of conductor depend on the voltage applied to elongated element, as well as on the placement of this point respect to the boundary of inversion of potential (BIP). BIP corresponds to geometric placing of points at the elongated element surface, where positive and negative potentials compensate each other. The transition through the BIP is accompanied by the potential sign inverse. As the potential difference at the edges of elongated element responds to the voltage applied, then the potential increase occur while distancing from BIP towards both edges of elongated element. Note that every point of the surface of elongated element in direction along its longitudinal axis is being ascribed by the certain value of electric potential. An existence of the difference of potentials between the surface areas connected by electrolyte is the enough condition for realization of the closed electrochemical circuits, which include electrodes segregated by ionic conductor, on a basis of elongated element. As the classic electrochemical electrode is an equipotent object, thus being determined by certain value of electric potential, it is possible to consider that some multi-electrode system is being realized at the surface of elongated element. This multi-electrode system is a sequence of the potential distribution along the elongated element. An existence of such kind of systems [5, 6] introduces determination in interpretation of the notion «electrode» and illustrates certain incorrectness of determining electrodes as the electronic conducting solid phase immersed into electrolyte. It is worth to consider an electrode as the semi-element consisting of a set of conducting phases included in series where oxidation or renovation reactions with participation of electrons are occurring [10].

The following experiment is a visual illustration of functioning of elongated element. The tips of the steel conductor (steel 08-kn) of the length of 70 cm and diameter 0.2 mm, immersed into the 0.5 mol/l$^3$ solution of NaCl, were connected to the voltage source terminals. According to the conditions of connection, it is possible to neglect the voltage decreasing at the circuit sections from the source terminals to the conductor waterline. At various values of voltage, which was supplied from the source, there were measured potentials at the conductor tips with a help of two silver-chloride reference electrodes. The results of measurements are exhibited in the plot (Fig. 3), which ordinate axis represents potential $E$ relative to the silver-chloride reference electrode, while abscissa axis presents voltage $U$, being applied to the conductor edges.

As it is clear from Fig. 3, at $U \leq 0.8$ V the function $E(U)$ is linear for both tips of conductor, which happens always due to the small values of the electrodes polarization. The voltage increase causes increasing of the electrode polarization and leads to appearance of nonlinear processes. Anodic section of elongated element becomes a place of not only dissolution of iron, but also of the oxygen separation and even the chlorine separation due to the further polarization. The reaction of renovation of protons occurs at the opposite (cathode) section of elongated element. The cathode reaction is being characterized by essentially higher electrode polarization than anodic one (Fig. 3). It is worth reminding that the difference of potentials between the conductor tips remains just that, which is equal to the voltage applied. Indeed, the load absent ($U = 0$) means that both tips of our conductor have the same stationary potential $E_s$. An appearance of the load arises the cathode $\Delta E^-$ and anodic $\Delta E^+$ polarization of the conductor edges, which is equal to the difference between the electrode potential at $U>0$ and the stationary potential $E_s$. The conductor edge, being connected with the negative terminal of the source, is being polarized in the cathode manner, while that connected with the positive terminal acts as anode. Our experimental data show that summation of the absolute values of $\Delta E^+$ and $\Delta E^-$ (Fig. 3) for any cases gives a value corresponding to that of abscissa of $U$.

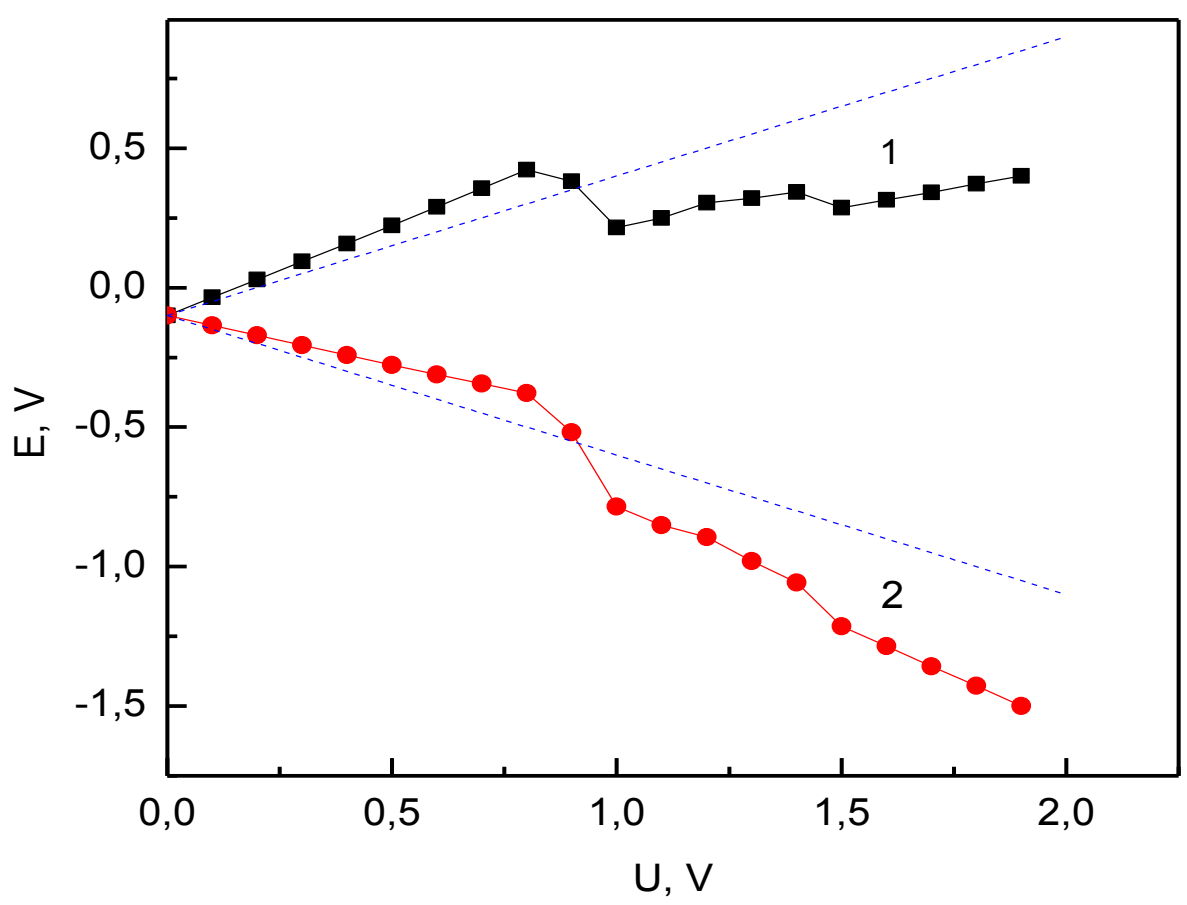

**Figure 3.** Dependence of potential $E$ of the anode (curve 1) and cathode (curve 2) polarized tips of the elongated element on applied voltage $U$ (with respect to the silver-chloride reference electrode). The dotted lines show dependence $E = \pm 0{,}5U + E_s$, where $E_s$ is stationary potential of the conductor tips at voltage $U = 0$.

Noteworthy the voltage increasing leads to the progressive asymmetry of potentials at the elongated element tips. The value of polarization of hetero-phase reactions is determined, in particular, by the nature of occurring processes, as well as by the substrate material. An asymmetry of function $E(U)$ may appear because of the different nature of reactions occurring at the conductor tips, as well as a sequence of the surface transformations due to the influence of products of these reactions. An asymmetry, which is being observed, can evidence about an existence of the potential difference between the polarized material surface layer, where electrochemical hetero-phase reactions occur, on one hand, and internal layer, being polarized as well, but isolated from chemical influences, on the other hand. Meanwhile chemical reactions do not happen inside elongated element, where direct contact with electrolyte is absent. Consequently, the potentials of edges of elongated element are determined as a result of symmetric distribution of applied voltage, and, so, have the values of $+U/2$ for a tip, being connected to the positive terminal of the power source, and $-U/2$ for the opposite tip. So, in absence of contact with electrolyte the dependence of potential of the elongated element tip, which is connected with the positive terminal of source, on the voltage applied is of the form $E = 0.5U + E_s$, while for an edge connected to the negative terminal such dependence is $E = -0.5U + E_s$ (Fig. 3.3). Assuming that the coordinate $x$=0 responds to the conductor edge being connected with the negative terminal of source, the longitudinal distribution of potential in the bulk elongated element of the length $l$ should be written down in a general form of $E = U(x/l - 0.5)$. Comparing experimental data and analytical dependences in Fig. 3, it is possible to determine the difference between potentials of internal and external layers of conductor. For instance, at the voltage of 750 mV the external layers of the anode and cathode tips of conductor are approximately 100 mV more positive than potentials of internal layers. At the same time, the voltage increase by two times up to 1500 mV leads to the negative shift of 400 mV of potential of external layer of anodic tip, as well as to the cathode tip negative shift of 300 mV with respect to potentials of internal layers. So, using the elongated element concept, it is easy to estimate the differences of potentials of the bulk and surface in any point of conductor as a function of the electrode polarization. These results should be very informative ones, in particular, for description of processes of adsorption and forming of the films of chemical compounds at the electrode surface [11-13].

Macroscopic elongated element under consideration has the round cross section of the constant diameter along the whole longitudinal axis. In general case the conductor cross section may have complex configuration with varying diameter. As for our case it is expected that metallic conductor connecting the counter-electrodes is a dendrite, so, the truncated cone may be considered in a role of approximating figure. This figure represents the peculiarity of configuration of dendrites, which diameter is almost always essentially smaller than that of its basis. Let us assume that the cone basis intersects the coordinate axis, coinciding with the axis of truncated cone, in a point $x = 0$, while dendrite top intersects this axis in a point $x = l_{el}$, where $l_{el}$ is the length of the elongated element in a form of truncated cone. At the monotonous and linear variation of the cone diameter $d_x$ along the cone axis it is possible to write in

$$d_x = d_1 + x/l_{el}\,(d_2 - d_1) \quad , \tag{1}$$

where $d_1$ and $d_2$ are diameters of the base and of the top of truncated cone correspondingly.

In electric field electric current flows through the body of dendrite, which is not surrounded by the ion-conducting medium. Meanwhile any surface electrochemical processes are absent. According to the Ohm's law and taking into account the equation (1), the resistance of dendrite section from 0 to $x$ can be represented in the following form:

$$R_x = \rho \frac{l_{el}}{S_x} = \frac{4\rho}{\pi} \int_0^x \frac{dx}{\left[ d_1 + \frac{x}{l_{el}} (d_2 - d_1) \right]^2} = \frac{4\rho x}{\pi d_1 \left[ d_1 + \frac{x}{l_{el}} (d_2 - d_1) \right]} , \tag{2}$$

where $S_x$ is the cross section of the cone with diameter $d_x$.

If voltage $U$ is applied to dendrite under consideration, so current $I$, flowing through this dendrite, is determined by the total resistance, which can be found from the equation (2) at $x = l_{el}$. Knowing the values of electric current and dendrite resistance at the section from 0 to $x$, it is possible to determine the value of decreasing of voltage $U_x$ at this section $U_x = IR_x$. Obviously the distribution of the voltage decreasing along dendrite axis is determined by the following relationship

$$U_x = \frac{U \delta x_r}{\left(1 + x_r \left(\delta - 1\right)\right)} , \tag{3}$$

where $\delta = d_2/d_1$ is the taper factor; $x_r = x/l_{el}$ is the normalized coordinate of the point at dendrite axis.

Saving an action of electric field, let us immerse our dendrite into the ion-conducting medium. This leads to appearance of the multi-electrode system at the sample surface with occurrence of cumulative electrochemical reactions. To estimate the distribution of the electrode polarization, one has to know the coordinate of BIP. The latter parameter is determined as coordinate of the point of intersection of the BIP plane with the main axis of elongated element, which is represented by a dendrite. An equation for normalized coordinate of BIP, which is noted by the letter $\alpha$, can be obtained from the expression (3), taking into account that the mutual compensation of the positive and negative potentials corresponds to the condition $U_x/U = 0.5$:

$$\alpha = \left(1 + \delta\right)^{-1} \tag{4}$$

At the inversion boundary the electrodes polarization of the multi-electrode system is absent by definition, hence, the cumulative electrochemical processes do not occur. Every certain coordinate $x_r \neq \alpha$ is in mutual correspondence with the certain potential of the polarized electrode of this system; therefore, polarization in any point of dendrite surface can be represented as

$$\Delta E(x_r) = U(x_r) - U(\alpha) = U(x_r) - U/2.$$

Taking into account the expressions (3) and (4), one can obtain

$$\Delta E = \frac{U}{2}\frac{(x_r - \alpha)}{\left(\alpha + (1-2\alpha)x_r\right)}. \tag{5}$$

The graphical presentation of the equation (5) in Fig. 4 illustrates visually the fact that approaching of the value of $\alpha$ to 1 causes increasing of nonlinearity of the polarization distribution along the elongated element. Thus, at $\alpha$ >0.95, as it is seen from Fig. 4, the most part of the voltage applied is falling in an area of dendrite surrounding the top of the latter.

Using the expression (4), one can conclude that the marginal case $\alpha \rightarrow 1$ takes place when diameter of dendrite top becomes very small, thus folding just into a point. As a result, the zone of occurrence of electrochemical reactions localizes at the top of truncated cone. An area of electrochemically active surface of elongated element forms at the section from dendrite top until the boundary of maximum of the cathode polarization. Let us note the point of intersection of the plane, which the boundary of maximum of the cathode polarization is placed in, with the elongated element axis as a coordinate of the latter boundary. The normalized value of coordinate of the boundary of maximum of the cathode polarization, which is equal to the ratio of relevant interval of dendrite axis to its general length, will be signed by the letter $\beta$.

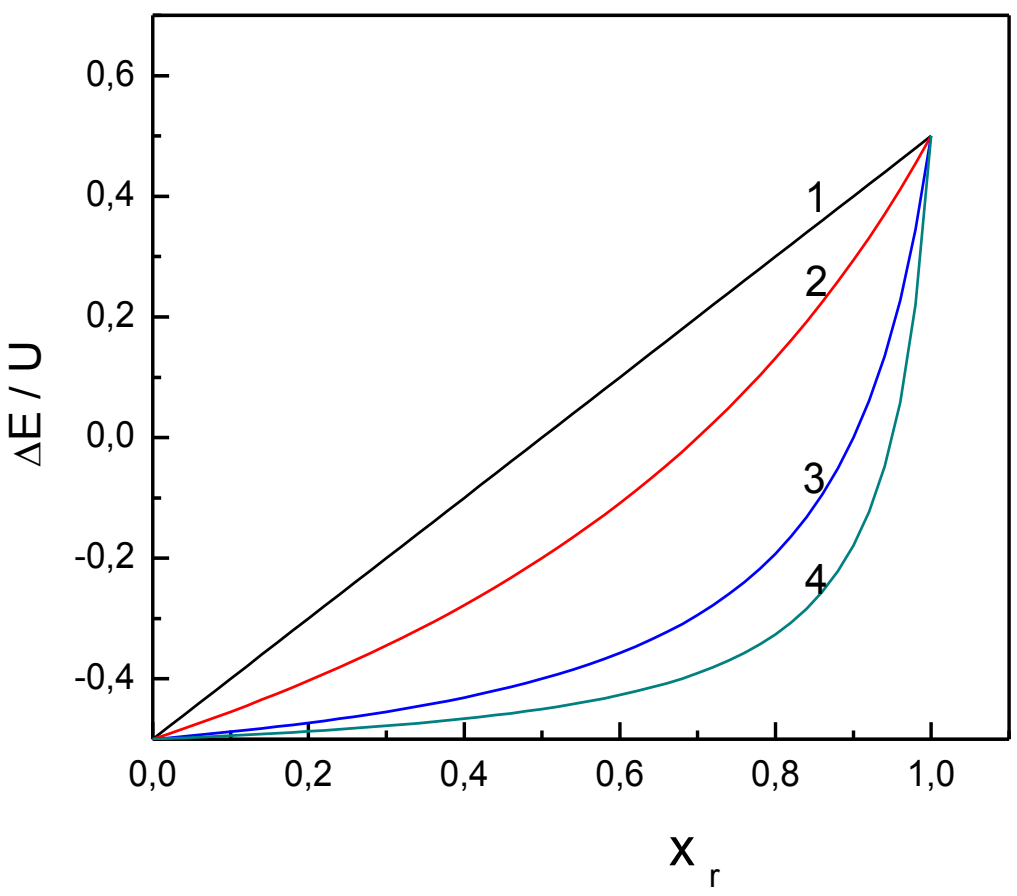

**Figure 4.** Distribution of polarization along the elongated element at various values of coordinate of the potential inversion boundary $\alpha$*:* 0.5 (curve 1); 0.7 (curve 2); 0.9 (curve 3); 0.95 (curve 4).

The coordinate $\beta$ can be determined, basing on the condition of conjugation of the anode and cathode electrochemical processes. As direction of dendrite growth is determined by polarity of the electrode system connection to the power supply source, so we may consider that the coordinate $x_r = 0$ is corresponded by the commutation point of elongated element with the negative terminal of the supply source, while the coordinate $x_r = 1$ just by the commutation point with the positive terminal. The coordinate $\beta$ divides our element into two sections: the section [0; $\beta$] is non-polarized one, serving as a conductor, which supplies electric current to the polarized section [$\beta$; 1], on the surface of which electrochemical processes are occurring. Let us assume that electrochemical reactions occur with participation of the particles of one kind and that the current density is connected linearly with the electrode polarization [7], so

$$j_-(x_r) = k_- \Delta E_-(x_r); \quad j_+(x_r) = k_+ \Delta E_+(x_r), \tag{6}$$

where $j_-(x_r)$ and $j_+(x_r)$ are densities of the cathode and anode currents respectively as the functions of coordinate at the longitudinal axis of elongated element; $\Delta E_-(x_r)$ and $\Delta E_+(x_r)$ are the cathode and anode polarization respectively; $k_-$ and $k_+$ are coefficients of proportionality. Taking into account the condition of conjugation of the cathode and anode processes, characterizing by equality of the anode and cathode currents, it is possible to determine the coordinate $\beta$ of the boundary of the cathode-polarized surface:

$$\int_{\beta}^{\alpha} j_-\left(x_r\right) dS\left(x_r\right) = \int_{\alpha}^{1} j_+\left(x_r\right) dS\left(x_r\right), \tag{7}$$

where $S(x_r) = \pi A d_1 \left[ (x_r + \frac{x_r^2(1-2\alpha)}{2\alpha} \right]$ is a square of the part of the side surface of elongated element, which is restricted by the plane intersecting an axis in the point $x_r$; $A = \sqrt{\frac{(d_1 - d_2)^2}{4} + l_{el}^{\ 2}}$ .

By means of solving the equations (7) one can obtain the following relationship

$$\beta = \alpha \left( 1 + \sqrt{\frac{k_+}{k_-}} \right) - \sqrt{\frac{k_+}{k_-}} . \qquad (8)$$

For a case of symmetry of the anode and cathode branches of the polarization characteristic of conjugated electrode reactions $k_+ = k_-$, the expression (8) means clearly that the coordinate $\beta$ is determined by the sizes of the basis and of the top of truncated cone:

$$\beta = \frac{d_1 - d_2}{d_1 + d_2} \qquad (9)$$

At rather small diameter of dendrite top ($d_2 \rightarrow 0$), creating the point contact together with the counter-electrode, the value $\beta$ approaches to 1, and all electrochemical transformations concentrate at dendrite surface, which surrounds directly the contact.

Basing on the obtained results, we have an opportunity of estimating the contribution of electrochemical current to the total current in elongated element. This simple estimation is reduced to determining of the ratio of the cathode or anode current flowing through the multi-electrode system, which is being formed at dendrite surface, to the total electronic current. If the voltage $U$ is applied to the elongated element, modeling a dendrite, then the value of electronic current $I_e$, which is being supplied to elongated element of the length $l_{el}$ from the negative terminal of source in absence of the concurrent channels of conductivity (hence, in absence of electrochemical reactions), is equal to $I_e = U/R$. Using the equation (2) for determining of dendrite resistance at $x = l_{el}$, one obtains

$$I_e = \frac{\pi U d_1 d_2}{4 \rho l_{el}} . \qquad (10)$$

While electric current is flowing through elongated element, the value $I_e$ remains unchanged at the section [0; $\beta$]. Meanwhile in the interval [$\beta$; $\alpha$] the concurrent conductivity channels through electrolyte appear, and electronic current decreases down just to the inversion point $\alpha$. This fact is caused by decreasing of the concentration of the current carriers due to participation of electrons in electrochemical reactions. The rate of the current decreasing is maximal in the cross section intersecting the axis of elongated element in a point $\beta$. The latter circumstance is caused by the maximal intensity of the cathode reaction. The rate of varying of electronic current further decreases monotonously, and becomes equal to zero in the inversion point. Noteworthy the concentration of carriers in the cross section, including the inversion point, reaches minimum, and the current value in this point is determined as a difference of electronic and total cathode current, which can be determined according to the left side of equation (7). The monotonous increase of electronic current occurs at the section [$\alpha$; 1] of prolonged element, and the value of electronic current in the cross section containing the point $x_r = 1$ is determined by equation (10). The character of changes of ionic current along the axis is just the same, therefore, during transition through BIP anodic current increases monotonously and reaches its maximum at dendrite top. The total anodic current $I_+$ is determined by the right side of equation (7) and has the following form:

$$I_+ = \pi k_+ \frac{U}{4} d_1 A \frac{(\alpha - 1)^2}{\alpha} . \qquad (11)$$

While dividing the expression (11) to the value of the total electronic current $I_e$ (10), one can determine, which part of the carriers $\theta$ participates in electrochemical process. Taking into account the relationship (4), we obtain the following result:

$$\theta = k_+ \rho l_{el} (1 - \alpha) \sqrt{\left( \frac{1}{2\alpha} - 1 \right)^2 + \left( \frac{l_{el}}{d_1} \right)^2} . \qquad (12)$$

Let us consider the limiting cases of varying of the value $\theta$ depending on geometric parameters of elongated element. Decreasing of diameter $d_2$ of dendrite top at unchanged relationship $l_{el} /d_1$, leads to decreasing of the value $\theta$. At the same time, at $l_{el} /d_1 \to 0$ the maximum of contribution of electrochemical current appears with the coordinate of extremum $\sqrt{(0,5)}$, corresponding to the value of $\delta=d_2/d_1$=0.414. General character of dependence of $\theta$ on coordinate of the inversion point is illustrated by Fig. 5. The obtained value of $\theta$ characterizes the quantity of carriers participating in electrochemical process, but does not reflect the latter process intensity. The degree of intensity of electrochemical processes is ascribed by the current density value. As electrochemical current flowing through the surface of elongated element is distributed along the axis, it is possible to estimate the average value of the current density. For this estimation it is enough to divide the value of anodic current (11), flowing in the multi-electrode system, to the square of the surface, through which this current flows. This square is equal to the side surface of dendrite from BIP to the top of the latter. So, the equation for average density of anodic current has the following form:

$$\hat{j}_+ = k_+ \frac{U}{2} \frac{(1+\delta)}{(3+\delta)} \tag{13}$$

At $k_+ = 1.3 \cdot 10^3\ \Omega^{-1}m^{-2}$ [14], it is easy to determine that at voltage of 1V the average density of anodic current is in a frame of 217-325 A/m$^2$. These values are at the level of the current density, being typical for the standard parameters at galvanostegic processes [14]. This fact evidences about ability of elongated element to ensure running of the electrode reactions with rather high intensity.

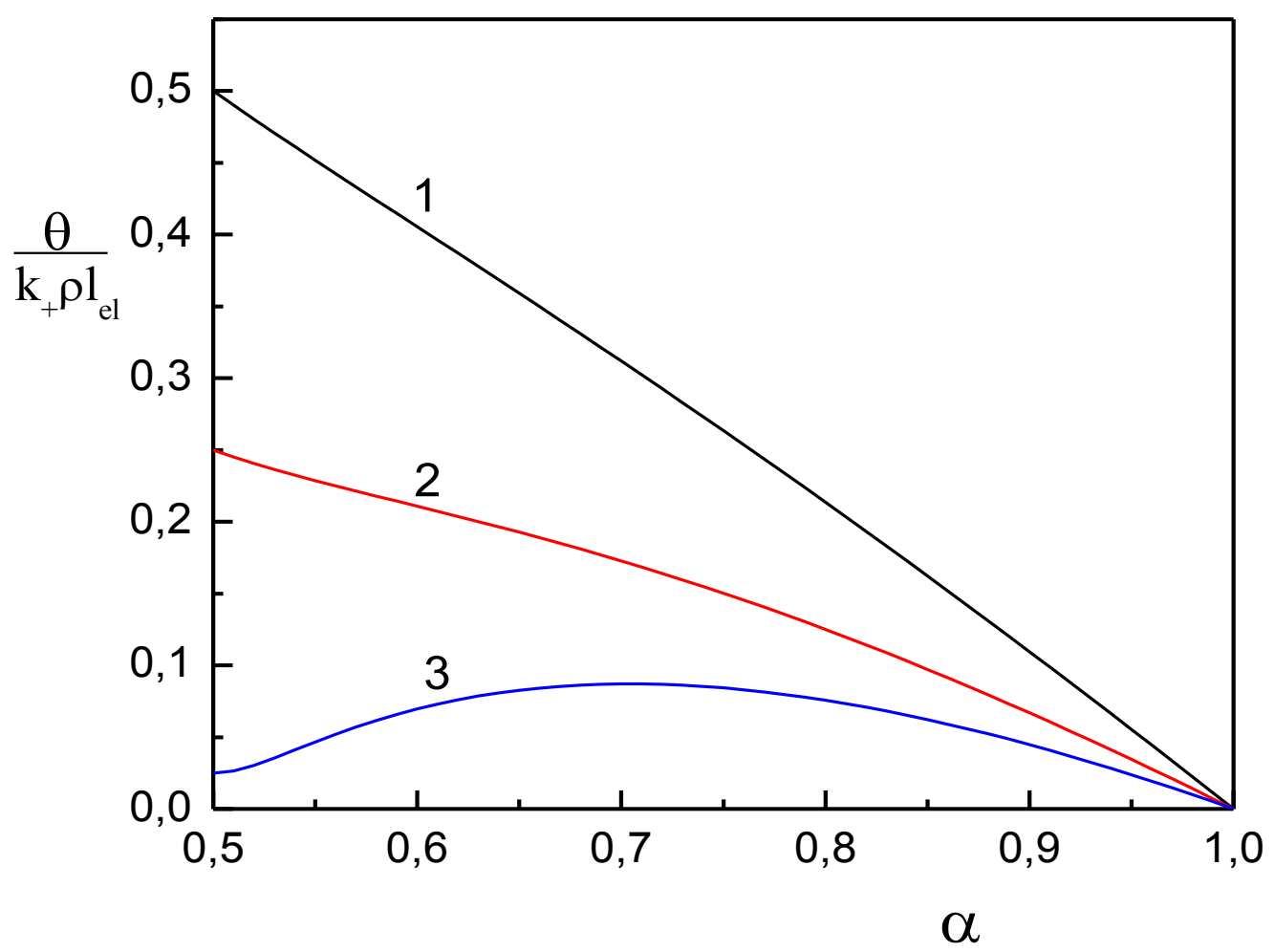

**Figure 5.** Variation of contribution of electrochemical current $\theta$ in the total current of elongated element depending on the coordinate of the potential inverse boundary $\alpha$. $l_{el}/d_1$ = 1 – dependence 1, $l_{el}/d_1$ = 0.5 – dependence 2, $l_{el}/d_1$ = 0.05 – dependence 3.

Let us dwell on the consequences, which should be expected due to the application of the elongated element concept to nano-objects. In the system under consideration and under condition $\alpha \to 1$ fulfilled at the place of the mutual touching of dendrite and counter-electrode, the nanostructure is being realized in a form of point contact [15, 16]. Indeed, expression (4) shows that $\delta \to 0$ for the latter case. As $\delta = d_2/d_1$, then at $d_1$ = const one can obtain $d_2 \to 0$, so, diameter of dendrite top shrinks to a point. So, at the place of touch of dendrite top to the surface of counter-electrode the point contact is being realized. Typical property of point contacts is that in a system "electrode-point contact-electrode" all the applied voltage is dropping just on this nanostructure [15, 16]. Taking into attention the distribution of polarization along the prolonged element (5) and respective analysis in Fig. 4, it is easy to find the correlation of the level of polarization of the surface of elongated element due to decrease of diameter $d_2$ with decreasing of the voltage in a contact. It is possible to conclude from this consideration that the ability of point contacts to concentrate the electric field in assemblage with the peculiar properties of elongated element causes unambiguously the localization of all the electrochemical transformations at the dendrite top, where contact is forming.

Besides, the point contact is a place of realization of the very high densities of electronic current, which vector is parallel to the axis of the contact channel [15, 16]. Note that the contact destroy does not happen due to the conduct away of the heat energy into so-called "banks-electrodes", which are essentially more massive metallic conductors delivering electric current. At the same time, the point contact in the ion-conducting medium does not have similar factor neutralizing concentration of energy of the cumulative movement of charges of electrochemical circuit. This circumstance leads to the fact that the point contact, which is being created in a process of dendrite growth, is destroyed quickly by the action of anodic polarization because of its own small size. As a result, electronic transport between electrodes is being stopped. Such kind of the contact destroying leads to the varying of the nature of the circuit, which current is flowing through. Instead of electronic conductor of the circuit it is being included more high resistive ionic conductor and two interface boundaries, one of which is a place of the anode process happening, while another is the cathode process place. Variation of dominating nature of conductivity in the system under consideration is being accompanied by the system resistance changing. This effect is observable experimentally due to the ensured growth of metallic dendrite in the inter-electrode space. Thus, during flow of direct current between copper needle and the plane copper plate being immersed into solution of blue vitriol, the hopping-like changes of the system resistance occur in the auto-oscillating fashion (Fig. 6).

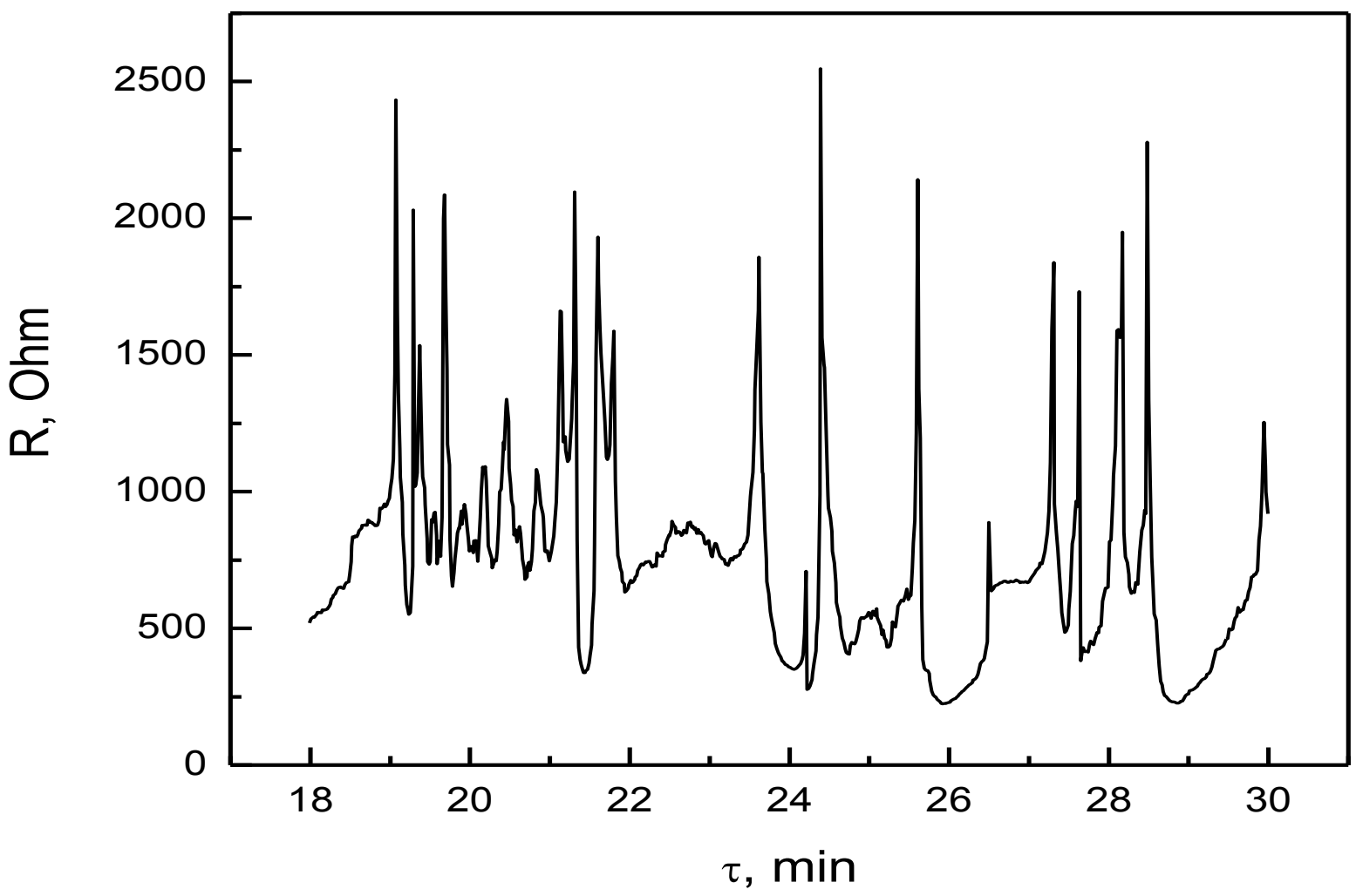

**Figure 6.** Fragment of chrono-resistogram of the system Cu│$CuSO_4$; $H_2O$ │Cu in the «needle-anvil» configuration. Current between electrodes is equal to 20 µA. *R* – resistance of the electrode system, $\tau$ – time.

The form and parameters of the chrono-resistogram reflect the properties of system and dynamics of changes of its states in the auto-oscillation process. General minimums of resistance, probably, correspond to the existence of elongated element in a form of dendrite connecting the counter-electrodes. Thus, the levels of minimums characterize the sizes of point contacts appearing in a place of dendrite top touching with the counter-electrode surface. The resistance maxima at chrono-resistograms correspond to the moments of destroying of the point contact because of anodic dissolution. In galvano-static conditions after destroying of contact all the current flows through electrolyte and through two interface boundaries «metal-electrolyte». Ions are playing the current carriers role in a clearance between dendrite top and counter-electrode. As ions are larger and, therefore, less mobile particles in comparison with electrons [7], that is one of the factors of the system resistance increasing. The intermediate local extremums of the chrono-resistogram can be connected with so-called dendrite branching, just an appearance and development of nucleus of metallic phase at the side surfaces of the main dendrite structure. Along with this phenomenon, some sections of growth, which were active earlier, become passive completely and do not participate in the charge transfer through the interface boundary. This circumstance can cause variation of the boundary resistance, as well as of general resistance of the system: an appearance of new areas of growth decreases polarizing resistance, and, on the contrary, passivity of previously active surface leads to increase of the polarizing resistance.

An appearance of dendrite point contact in a process of auto-oscillations is being accompanied by a series of transitional processes. The nature and character of the latter processes depend on the relation between the voltage applied and the system resistance. At the moment of touching of dendrite top to the counter-electrode the channel of electronic conductivity switches-in. In potentiostatic regime ($U$ = const) the latter circumstance leads to the drastic increase of current flowing through the system. As the electrode polarization stays unchanged, so, ionic current may flow concurrently to electronic one, causing an existence of electrochemical processes. Meanwhile in galvanostatic regime ($I$ = const), just on the contrary, the moment of contacting of dendrite to the counter-electrode is connected with the drastic decrease of the electrode reactions speed because of the polarization reducing. If the voltage drop on the dendrite exceeds the minimal voltage, serving for occurrence of the cumulative electrode reactions, than electrochemical current flows concurrently to electronic one similarly to the situation of potentiostatic regime. However if the condition mentioned is not fulfilled, electrochemical processes beyond the contact interrupt, and all the current flows through the dendrite. The equivalent electric schematic for this case (Fig. 2) simplifies considerably, having a form of the active resistance equal to dendrite one. Our experiments have shown that the auto-oscillatory effect is being observed mostly due to galvanostatic regime. While supporting the constant current, the latter regime secures the high concentration of the power lines of electric field at dendritic top, being created after electrochemical dissolution of the conductivity channel. Just the latter circumstance is ensuring the fulfillment of the necessary conditions for forming of the new dendritic contact, and, hence, for supporting of a process of periodic commutation of the needle and counter-electrode.

One of the models of the point contact, which is being used in practice widely, is a model of the constant cross section channel [15, 16]. Such kind of the contact geometry allows using the above expressions for elongated element, as well as estimating the parameters of electrochemical processes, which can happen in an area of the point-contact nanostructure. Electric resistance of a contact in a form of the long channel $L$ in the pure limit of the current flow, satisfying the condition $l_i>>L>>d$, is determined by the following relationship [15, 16]:

$$R_0 = 16\rho l/3\pi d^2, \qquad (14)$$

where $\rho$ is specific electric resistance of material, $l$ is the mean free path of electrons, $d$ is the channel diameter, and $l_i$ is the elastic length of the electron path, taking into account the scattering on impurities. For a case of dirty limit ($l_i<<d<<L$) the formula for electric resistance of the contact with the channel length $L$ corresponds to the Ohm's law and is of the form $R_0 = 4\rho L/\pi d^2$ [17, 18]. At room temperature for a case of contacts created by means of traditional methods more frequently is being realized an intermediate or dirty limit of the current flow. At the same time, as the electron mean free path in copper at room temperature may reach several hundred of Ängström [19], realization of the pure or intermediate limit of the current flow during formation of dendrite contacts is real enough. If diameter and length of the contact channel are considerably smaller or of an order of the electron mean free path, than practically all the applied voltage drops at the point contact [15, 16]. This peculiarity allows considering the point contact described by the channel model as elongated element of the length $l_{el} = L \approx l$ with the constant diameter $d$. It is possible to estimate the sizes of such contact with a help of the formula (14), knowing its resistance $R$ and assuming $R = R_0$.

Dendrite point contact appears in a place of touching of the counter-electrode surface by the top of a dendrite growing in electric field. According to equation (4) for a contact in a model of the constant cross section channel, the inversion boundary coordinate is $\alpha = 0.5$, situating just at the distance of $L/2$ from a point of touching of the surface by dendrite top. Let us insert the following parameters of the contact $d = d_1 = d_2$; $\alpha = 0.5$; $L = l_{el}$ in the expression (12) determining a share of the current carriers, which participate in electrochemical process. The resulting formula is as follows

$$\theta = \frac{k_+ \rho L^2}{2d}. \qquad (15)$$

Let us estimate numerical value of $\theta$ for the copper dendrite contact of diameter $d$=$10^{-9}$ m, using the known values of parameters present in the formula (15): $k_+ = 1.3\cdot 10^3\ \Omega^{-1}m^{-2}$ [14], $\rho = 1.6\cdot 10^{-8}\ \Omega\cdot m$ [20]. Assuming realization of the intermediate limit of the current flow, one may consider that $L= l = 10^{-8}$ m [19]. Inserting the listed values of parameters, we obtain the estimation of a factor $\theta \approx 1.0\cdot 10^{-12}$.

So, one can see that electrochemical current forms quite slight share of the electronic one, just very small, taking into account the experiment conditions during observation of the commuting process. At the same time, as it is possible to see from the chrono-resistogram in Fig. 6, just such minor current destroys the point contact during the time period about $\tau_{ex}$=1 s. It allows dimensions evaluation of structures under transformation and one can show that these dimensions are really very small. Using Faraday law $m = q\ I_+\tau_{ex}$, where $m$ is the mass of metal, which is being anode-dissolved due to action of electric current $I_+$ during a period of time $\tau_{ex}$ ($q$ is electrochemical equivalent), it is possible to obtain the relationship for volume $V$ of destroying share of the dendrite:

$$V \approx \frac{\theta I_e \tau_{ex} q}{\gamma}, \qquad (16)$$

where $\gamma$ is the metal density. For copper dendrite at $q = 3.32\cdot 10^{-7}$ kg/C and $\gamma = 9\cdot 10^3$ kg/m$^3$, $V \approx 10^{-27}$ m$^3$. This estimation corresponds to characteristic size about 1nm for the structure destroying and coincides with the channel diameter of point contact calculated from formula (14) at $R_0 \approx 300\ \Omega$. Hence, supplying electric current to dendrite point contact in the axis direction of the latter causes the flow of rather small anodic currents in the commutation zone of the conductivity channel with the bulk conductor. Nevertheless, due to the small sizes of the structure studied these currents manifest themselves as strong enough for destroying of the top of dendrite in a zone of the nanosized contact.

All the procedures of electrochemical synthesis of nanostructures, which are known up to date, are based on the usage of the classical architecture of the electrode systems [12, 21-24], causing the necessity of the technically complex control of the super-small currents and polarizations. In present work it is suggested the concept of continual multi-electrode system being realized on the elongated electrically conducting element immersed into electrolyte. The elongated element in present configuration works efficiently in the ranges of currents 1-100 μA and voltages 1-1000 mV. Conjugation of such kind of system with the point-contact structure creates the unique electrode architecture and permits realizing of the auto-oscillation process of electrochemical commutation at atomic scale. The point contacts, being created during the latter process, can find their practical applications due to the enhanced gas sensitivity [25]. The latter advantageous property causes the opportunities of usage of the obtained data for optimization of metrological parameters of analytic devices of the new type.

We acknowledge V. Gudimenko for fruitful discussions and assistance in the treatment of results. This work was supported by the Science and Technology Centre in Ukraine (STCU), and the National Academy of Sciences of Ukraine.

## References.